\documentclass[aps,prl,twocolumn,superscriptaddress]{revtex4}
\usepackage{amssymb,amsbsy,graphicx,xcolor,times,subfigure,marvosym,color,bm,multirow,epstopdf}
\usepackage[latin1]{inputenc}
\begin{document}

\title{Probing spinon nodal structures in three-dimensional Kitaev spin liquids}

\author{G\'abor B. Hal\'asz}
\affiliation{Kavli Institute for Theoretical Physics, University of
California, Santa Barbara, CA 93106, USA}

\author{Brent Perreault}
\affiliation{School of Physics and Astronomy, University of
Minnesota, Minneapolis, MN 55455, USA}

\author{Natalia B. Perkins}
\affiliation{School of Physics and Astronomy, University of
Minnesota, Minneapolis, MN 55455, USA}

%%%%%%%%%%%%%%%%%%%%%%%%%%%%%%%%%%%%%%%%%%%%%%%%%

\begin{abstract}

We propose that resonant inelastic X-ray scattering (RIXS) is an
effective probe of the fractionalized excitations in
three-dimensional (3D) Kitaev spin liquids. While the
non-spin-conserving RIXS responses are dominated by the gauge-flux
excitations and reproduce the inelastic-neutron-scattering response,
the spin-conserving (SC) RIXS response picks up the Majorana-fermion
excitations and detects whether they are gapless at Weyl points,
nodal lines, or Fermi surfaces. As a signature of symmetry
fractionalization, the SC RIXS response is suppressed around the
$\Gamma$ point. On a technical level, we calculate the exact SC RIXS
responses of the Kitaev models on the hyperhoneycomb,
stripyhoneycomb, hyperhexagon, and hyperoctagon lattices, arguing
that our main results also apply to generic 3D Kitaev spin liquids
beyond these exactly solvable models.

\end{abstract}

%%%%%%%%%%%%%%%%%%%%%%%%%%%%%%%%%%%%%%%%%%%%%%%%%

\maketitle

%%%%%%%%%%%%%%%%%%%%%%%%%%%%%%%%%%%%%%%%%%%%%%%%%

Quantum spin liquids (QSLs) are exotic and entirely quantum phases
of matter \cite{Balents-2010, Savary-2016} hosting a remarkable set
of emergent phenomena, including long-range entanglement,
topological ground-state degeneracy, and fractionalized anyonic
excitations. The Kitaev spin liquid (KSL) on the honeycomb lattice
\cite{Kitaev-2006} and its generalizations on tricoordinated
three-dimensional (3D) lattices \cite{Mandal-2009, Hermanns-2014,
Hermanns-2015, O'Brien-2016, Hermanns-2017} are quintessential
examples of such QSL phases. Importantly, recent years have seen
much progress in identifying a large number of candidate materials
for realizing these KSL phases \cite{Jackeli-2009, Chaloupka-2010,
Chaloupka-2013, Trebst-2017}, such as the honeycomb iridates
Na$_2$IrO$_3$ \cite{Singh-2010, Liu-2011, Singh-2012, Choi-2012,
Ye-2012, Comin-2012} and $\alpha$-Li$_2$IrO$_3$
\cite{Williams-2016}, the honeycomb ruthenium chloride
$\alpha$-RuCl$_3$ \cite{Plumb-2014, Sandilands-2015, Sears-2015,
Majumder-2015, Johnson-2015, Sandilands-2016, Banerjee-2016}, and
the 3D harmonic-honeycomb iridates $\beta$- and
$\gamma$-Li$_2$IrO$_3$ \cite{Modic-2014, Biffin-2014a, Biffin-2014b,
Takayama-2015}.

From a theoretical point of view, KSLs are particularly appealing
because each of them has an exactly solvable limit governed by a
Kitaev model \cite{Kitaev-2006}. In general, the Kitaev model is
defined on a tricoordinated lattice with $S = 1/2$ spins
$\sigma_{\mathbf{r}}^{x,y,z}$ at the sites $\mathbf{r}$, which are
coupled to their neighbors via bond-dependent Ising interactions.
The Hamiltonian reads
\begin{equation}
H = -J_x \sum_{\langle \mathbf{r}, \mathbf{r}' \rangle_x}
\sigma_{\mathbf{r}}^x \sigma_{\mathbf{r}'}^x - J_y \sum_{\langle
\mathbf{r}, \mathbf{r}' \rangle_y} \sigma_{\mathbf{r}}^y
\sigma_{\mathbf{r}'}^y - J_z \sum_{\langle \mathbf{r}, \mathbf{r}'
\rangle_z} \sigma_{\mathbf{r}}^z \sigma_{\mathbf{r}'}^z,
\label{eq-H-1}
\end{equation}
where $J_{x,y,z}$ are the coupling constants for the three types of
bonds $x$, $y$, and $z$. Remarkably, this model is exactly solvable
whenever there is precisely one bond of each type around each site
of the tricoordinated lattice.

\begin{figure*}
\includegraphics[width=1.95\columnwidth]{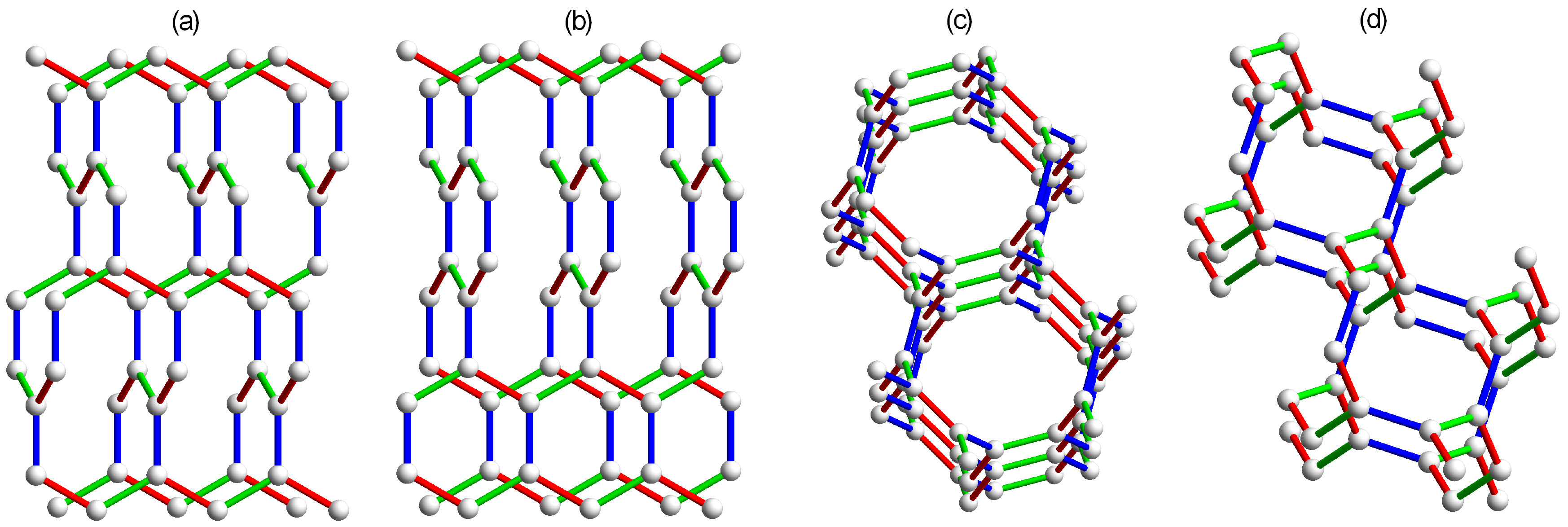}
\caption{Tricoordinated 3D lattices of the Kitaev models considered
in this work: (a) hyperhoneycomb, (b) stripyhoneycomb, (c)
hyperhexagon, and (d) hyperoctagon lattices. Different bond types
$x$, $y$, and $z$ are marked by red, green, and blue, respectively.}
\label{fig-1}
\end{figure*}

These exactly solvable Kitaev models have been defined on a wide
range of tricoordinated 3D lattices \cite{Mandal-2009,
Hermanns-2014, Hermanns-2015, O'Brien-2016, Hermanns-2017},
including the hyperhoneycomb, stripyhoneycomb, hyperhexagon, and
hyperoctagon lattices (see Fig.~\ref{fig-1}). In the experimentally
relevant isotropic regime ($J_x \approx J_y \approx J_z$), the
ground state is a gapless $\mathbb{Z}_2$ QSL, while the
(fractionalized) excitations are gapless Majorana fermions and
gapped $\mathbb{Z}_2$ gauge fluxes. Importantly, the Majorana
fermions (spinons) exhibit a rich variety of nodal structures due to
the different (projective) ways symmetries can act on them
\cite{Hermanns-2014, Hermanns-2015, O'Brien-2016}. Indeed, they are
gapless along nodal lines for the hyperhoneycomb and the
stripyhoneycomb models \cite{Mandal-2009}, on Fermi surfaces for the
hyperoctagon model \cite{Hermanns-2014}, and at Weyl points for the
hyperhexagon model \cite{O'Brien-2016}.

From an experimental point of view, however, it is difficult to
identify and characterize QSLs due to the lack of any local order
parameters that could be used as "smoking-gun" signatures. In recent
years, a remarkable theoretical and experimental progress has been
achieved in understanding that fractionalization is one of the most
promising hallmarks of a QSL. Indeed, it has been demonstrated that
fractionalized excitations, which are Majorana fermions and
$\mathbb{Z}_2$ gauge fluxes for KSLs, can be probed by conventional
spectroscopic techniques, such as inelastic neutron scattering (INS)
\cite{Banerjee-2016, Knolle-2014a, Knolle-2015, Smith-2015,
Smith-2016}, Raman scattering with visible light
\cite{Sandilands-2015, Sandilands-2016, Knolle-2014b,
Perreault-2015, Perreault-2016a, Perreault-2016b, Glamazda-2016},
and resonant inelastic X-ray scattering (RIXS) \cite{Ko-2011,
Savary-2015, Halasz-2016}.

In this Letter, we propose that RIXS is an effective probe of the
spinon (semi)metals realized in 3D KSLs. Calculating the exact RIXS
responses of four different 3D Kitaev models (see lattices in
Fig.~\ref{fig-1}), we demonstrate that nodal lines, Weyl points, and
Fermi surfaces of Majorana fermions leave distinct characteristic
fingerprints in the spin-conserving (SC) RIXS response. For the
hyperhoneycomb and the stripyhoneycomb models, corresponding to
$\beta$- and $\gamma$-Li$_2$IrO$_3$, the SC RIXS response is gapless
within particular high-symmetry planes but not at a generic point of
the Brillouin zone. In contrast, for the hyperhexagon model, it is
gapless at particular points only, while for the hyperoctagon model,
it is gapless in almost the entire Brillouin zone. Also, the SC RIXS
response is found to be strongly suppressed around the $\Gamma$
point for all four models as a result of symmetries acting
projectively on the Majorana fermions. We argue that our results
apply to generic KSLs and not only to the pure Kitaev models.

\emph{General RIXS formalism.---}Motivated by the available
candidate materials ($\beta$- and $\gamma$-Li$_2$IrO$_3$), we
calculate the RIXS responses for the $L_3$ edge of the Ir$^{4+}$ ion
which is in the $5d^5$ configuration \cite{Ament-2011a, Kim-2017}.
However, our results are also expected to be valid for other RIXS
edges and for other potential $d^5$ candidate materials
\cite{Halasz-2016}. During RIXS, an incoming photon is absorbed and
excites a $2p$ core electron into the $5d$ valence shell, which then
decays back into the $2p$ core hole and emits an outgoing photon
\cite{Ament-2011b}. The low-energy physics of the $5d$ valence shell
at each Ir$^{4+}$ ion is governed by a $J = 1/2$ Kramers doublet in
the $t_{2g}$ orbitals, and we assume that the low-energy Hamiltonian
acting on these Kramers doublets is the Kitaev Hamiltonian in
Eq.~(\ref{eq-H-1}). In terms of the corresponding Kitaev model, the
$5d^6$ configuration in the intermediate state is then described as
a non-magnetic vacancy \cite{Willans-2010, Willans-2011,
Halasz-2014, Sreejith-2016}.

The initial and the final states of RIXS are $| 0 \rangle \otimes |
\mathbf{Q}, \bm{\epsilon} \rangle$ and $| m \rangle \otimes |
\mathbf{Q}', \bm{\epsilon}' \rangle$, respectively, where $| 0
\rangle$ is the ground state of the Kitaev model, $| m \rangle$ is a
generic eigenstate with energy $E_m$ with respect to $| 0 \rangle$,
while $\mathbf{Q}$ ($\mathbf{Q}'$) is the momentum and
$\bm{\epsilon}$ ($\bm{\epsilon}'$) is the polarization of the
incoming (outgoing) photon. During RIXS, an energy $\omega = c \, \{
|\mathbf{Q}| - |\mathbf{Q}'| \} = E_m$ and a momentum $\mathbf{q} =
\mathbf{Q} - \mathbf{Q}'$ is transferred from the scattered photon
to the KSL. Summing over all final states $| m \rangle$, the total
RIXS intensity is then $I (\omega, \mathbf{q}) = \sum_m |A (m,
\mathbf{q})|^2 \, \delta (\omega - E_m)$, where $A (m, \mathbf{q})$
are the individual RIXS amplitudes.

Since RIXS has four fundamental channels \cite{Halasz-2016}, each
RIXS amplitude takes the form $A (m, \mathbf{q}) = \sum_{\eta}
P_{\eta} A_{\eta} (m, \mathbf{q})$, where $P_{\eta}$ are
polarization factors depending on $\bm{\epsilon}$ and
$\bm{\epsilon}'$ \cite{Ament-2011a}, while $A_{\eta} (m,
\mathbf{q})$ are single-channel RIXS amplitudes corresponding to the
four fundamental channels. In the SC channel labeled by $\eta = 0$,
the spin of the $5d$ valence shell does not change during RIXS,
while in the three non-spin-conserving (NSC) channels labeled by
$\eta = x,y,z$, the same spin is rotated by $\pi$ around the $x,y,z$
axes, respectively.

The single-channel RIXS amplitudes $A_{\eta} (m, \mathbf{q})$ are
given by the Kramers$-$Heisenberg formula \cite{Ament-2011b}. In the
experimentally relevant fast-collision regime, where the core-hole
decay rate $\Gamma$ is much larger than the Kitaev coupling
constants $J_{x,y,z}$ (e.g., for the iridates: $\Gamma / J_{x,y,z}
\sim 100$) \cite{Clancy-2012, Katukuri-2014}, these RIXS amplitudes
take the lowest-order form \cite{Halasz-2016}
\begin{eqnarray}
A_{\eta} (m, \mathbf{q}) &\propto& \sum_{\mathbf{r}} e^{i \mathbf{q}
\cdot \mathbf{r}} \langle m | \sigma_{\mathbf{r}}^{\eta} \bigg[ 1 -
\frac{i \tilde{H} (\mathbf{r})} {\Gamma} \bigg] | 0
\rangle \label{eq-A-1} \\
&=& \sum_{\mathbf{r}} e^{i \mathbf{q} \cdot \mathbf{r}} \langle m |
\sigma_{\mathbf{r}}^{\eta} \bigg[ 1 - \frac{i}{\Gamma} \sum_{\kappa
= x,y,z} J_{\kappa}^{\phantom{\kappa}} \sigma_{\mathbf{r}}^{\kappa}
\sigma_{\kappa (\mathbf{r})}^{\kappa} \bigg] | 0 \rangle, \nonumber
\end{eqnarray}
where $\tilde{H} (\mathbf{r}) = H + \sum_{\kappa}
J_{\kappa}^{\phantom{\kappa}} \sigma_{\mathbf{r}}^{\kappa}
\sigma_{\kappa (\mathbf{r})}^{\kappa}$ is the Hamiltonian of the
Kitaev model with a single vacancy at site $\mathbf{r}$. The spin at
site $\mathbf{r}$ is effectively removed from the model by being
decoupled from its neighbors at sites $\kappa (\mathbf{r})$
\cite{Halasz-2014}. Note also that $\sigma_{\mathbf{r}}^0$ is the
identity operator and that we demand $H | 0 \rangle = 0$ by adding a
trivial constant term to $H$ in Eq.~(\ref{eq-H-1}).

For the NSC channels, the RIXS amplitudes in Eq.~(\ref{eq-A-1})
reduce to spin-polarized INS amplitudes $\sum_{\mathbf{r}} e^{i
\mathbf{q} \cdot \mathbf{r}} \langle m | \sigma_{\mathbf{r}}^{x,y,z}
| 0 \rangle$ in the limit of $\Gamma \rightarrow \infty$. In the
physically relevant regime, the three NSC RIXS responses thus
reproduce the respective components of the dynamical spin structure
factor studied in Refs.~\cite{Smith-2015} and \cite{Smith-2016}.
Indeed, since the NSC channels involve flux creation, the
corresponding responses exhibit an overall flux gap and little
momentum dispersion \cite{Halasz-2016}.

For the SC channel, however, taking the limit of $\Gamma \rightarrow
\infty$ in Eq.~(\ref{eq-A-1}) gives a trivial amplitude
$\sum_{\mathbf{r}} e^{i \mathbf{q} \cdot \mathbf{r}} \langle m | 0
\rangle$ that corresponds to a purely elastic process. The
lowest-order inelastic process is then captured by the second term
in Eq.~(\ref{eq-A-1}), and the corresponding RIXS amplitude can be
calculated via the exact solution of the Kitaev model
\cite{Kitaev-2006}. Furthermore, since the SC channel creates no
fluxes, the entire calculation is restricted to the ground-state
flux sector of the model.

\emph{Spinon band structures.---}As a first step of our calculation,
we describe the fermion (spinon) band structures of the four Kitaev
models. Using the Kitaev fermionization
$\sigma_{\mathbf{r}}^{\kappa} = i b_{\mathbf{r}}^{\kappa}
c_{\mathbf{r}}^{\phantom{\kappa}}$ with $\kappa = x,y,z$, the
Hamiltonian in Eq.~(\ref{eq-H-1}) becomes
\begin{equation}
H = \sum_{\kappa} \sum_{\langle \mathbf{r}, \mathbf{r}'
\rangle_{\kappa}} i J_{\kappa}^{\phantom{\kappa}} u_{\mathbf{r},
\mathbf{r}'}^{\kappa} c_{\mathbf{r}}^{\phantom{\kappa}}
c_{\mathbf{r}'}^{\phantom{\kappa}} = \frac{1}{2} \sum_{\mathbf{r},
\mathbf{r}'} \mathcal{H}_{\mathbf{r}, \mathbf{r}'} c_{\mathbf{r}}
c_{\mathbf{r}'}, \label{eq-H-2}
\end{equation}
where $u_{\mathbf{r}, \mathbf{r}'}^{\kappa} \equiv i
b_{\mathbf{r}}^{\kappa} b_{\mathbf{r}'}^{\kappa} = \pm 1$ in the
ground-state flux sector, while $\mathcal{H}_{\mathbf{r},
\mathbf{r}'} = i J_{\kappa}^{\phantom{\kappa}} u_{\mathbf{r},
\mathbf{r}'}^{\kappa}$ if $\mathbf{r}$ and $\mathbf{r}'$ are
neighboring sites connected by a $\kappa$ bond and
$\mathcal{H}_{\mathbf{r}, \mathbf{r}'} = 0$ otherwise. It is known
that the ground state of the hyperhexagon model has a $\pi$ flux at
each elementary loop \cite{O'Brien-2016, Lieb-1994}, while we assume
that the ground states of the remaining three models are flux free.
This choice is consistent with numerical results for the
hyperhoneycomb and the hyperoctagon models \cite{Mandal-2009,
O'Brien-2016}, while it is merely a simplification for the
stripyhoneycomb model \cite{Footnote-1}.

The quadratic fermion Hamiltonian in Eq.~(\ref{eq-H-2}) can be
diagonalized via a standard procedure. Since the lattice of each
Kitaev model has $n$ sites per unit cell ($\nu = 1, 2, \ldots, n$),
the resulting band structure has $n$ fermion bands ($\mu = 1, 2,
\ldots, n$), where $n = 4$ for the hyperhoneycomb and the
hyperoctagon models, $n = 6$ for the hyperhexagon model, and $n = 8$
for the stripyhoneycomb model. For a lattice of $N$ unit cells, the
fermion with band index $\mu$ and momentum $\mathbf{k}$ takes the
form
\begin{equation}
\psi_{\mathbf{k}, \mu}^{\dag} = \frac{1} {\sqrt{2N}} \sum_{\nu =
1}^{n} \big( \mathcal{W}_{\mathbf{k}}^{\phantom{\dag}} \big)_{\nu
\mu} \sum_{\mathbf{r} \in \nu} c_{\mathbf{r}}^{\phantom{\dag}} \,
e^{i \mathbf{k} \cdot \mathbf{r}}, \label{eq-psi}
\end{equation}
while the corresponding fermion energy is $\varepsilon_{\mathbf{k},
\mu} = 2 (\mathcal{E}_{\mathbf{k}})_{\mu \mu}$, where
$\hat{\mathcal{H}}_{\mathbf{k}}^{\phantom{\dag}} =
\mathcal{W}_{\mathbf{k}}^{\phantom{\dag}} \cdot
\mathcal{E}_{\mathbf{k}}^{\phantom{\dag}} \cdot
\mathcal{W}_{\mathbf{k}}^{\dag}$ is the (unitary) eigendecomposition
of the Hermitian matrix
$\hat{\mathcal{H}}_{\mathbf{k}}^{\phantom{\dag}}$ with elements
\begin{equation}
(\hat{\mathcal{H}}_{\mathbf{k}})_{\nu \nu'} = \frac{1} {N}
\sum_{\mathbf{r} \in \nu} \sum_{\mathbf{r}' \in \nu'}
\mathcal{H}_{\mathbf{r}, \mathbf{r}'} \, e^{i \mathbf{k} \cdot
(\mathbf{r}' - \mathbf{r})}. \label{eq-H-3}
\end{equation}
Note that only the fermions $\psi_{\mathbf{k}, \mu}^{\dag}$ with
energies $\varepsilon_{\mathbf{k}, \mu}^{\phantom{\dag}} > 0$ are
physical due to the particle-hole redundancy
$\hat{\mathcal{H}}_{-\mathbf{k}}^{\phantom{\dag}} =
-\hat{\mathcal{H}}_{\mathbf{k}}^{*}$ which implies
$\psi_{-\mathbf{k}, \mu}^{\phantom{\dag}} = \psi_{\mathbf{k},
\mu}^{\dag}$ and $\varepsilon_{-\mathbf{k}, \mu}^{\phantom{\dag}} =
-\varepsilon_{\mathbf{k}, \mu}^{\phantom{\dag}}$. In terms of these
fermions, the Hamiltonian in Eq.~(\ref{eq-H-2}) is then
\begin{equation}
H = \sum_{\mathbf{k}} \sum_{\mu = 1}^{n} \varepsilon_{\mathbf{k},
\mu}^{\phantom{\dag}} \left[ \psi_{\mathbf{k}, \mu}^{\dag}
\psi_{\mathbf{k}, \mu}^{\phantom{\dag}} - \frac{1}{2} \right] \Theta
(\varepsilon_{\mathbf{k}, \mu}), \label{eq-H-4}
\end{equation}
where the Heaviside step function $\Theta (x) = \int_{-\infty}^x d
\tilde{x} \, \delta (\tilde{x})$ restricts the sum to physical
fermions.

At the isotropic point ($J_{x,y,z} = J_0$) of each Kitaev model,
there are gapless nodes in the band structure characterized by
$\varepsilon_{\mathbf{k}, \mu} = 0$. The structure of these nodes is
determined by how inversion and time-reversal symmetries act on the
fermions $\psi_{\mathbf{k}, \mu}^{\dag}$ \cite{Hermanns-2014,
Hermanns-2015, O'Brien-2016}. If time reversal is supplemented with
a momentum shift $\mathbf{k} \rightarrow \mathbf{k} + \mathbf{k}_0$,
the fermions are gapless at Weyl points in the presence of inversion
symmetry (hyperhexagon model) and on Fermi surfaces in the absence
of inversion symmetry (hyperoctagon model). If there is no momentum
shift associated with time reversal, the fermions are gapless along
nodal lines (hyper- and stripyhoneycomb models). For each model, the
matrix $\hat{\mathcal{H}}_{\mathbf{k}}$ and the band structure
$\varepsilon_{\mathbf{k}, \mu}$ are presented in the Supplementary
Material (SM) \cite{SM}.

\emph{SC RIXS responses.---}We are now ready to calculate the SC
RIXS responses of the four Kitaev models. Concentrating on the
second term of Eq.~(\ref{eq-A-1}) and using the Kitaev
fermionization, the lowest-order SC RIXS amplitudes are
\begin{equation}
A_0 (m, \mathbf{q}) \propto \sum_{\mathbf{r}, \mathbf{r}'} e^{i
\mathbf{q} \cdot \mathbf{r}} \, \mathcal{H}_{\mathbf{r},
\mathbf{r}'} \langle m | c_{\mathbf{r}} c_{\mathbf{r}'} | 0 \rangle.
\label{eq-A-2}
\end{equation}
For the inelastic processes $| m \rangle \neq | 0 \rangle$, the
final state $| m \rangle$ contains two fermions $\psi_{\mathbf{k},
\mu}^{\dag}$ and $\psi_{\mathbf{q} - \mathbf{k}, \mu'}^{\dag}$ with
a total momentum $\mathbf{q}$ and a total energy $E_m =
\varepsilon_{\mathbf{k}, \mu} + \varepsilon_{\mathbf{q} -
\mathbf{k}, \mu'}$. The lowest-order SC RIXS intensity of each
Kitaev model is then
\begin{eqnarray}
I_0 (\omega, \mathbf{q}) &\propto& \sum_{\mathbf{k}, \mu, \mu'}
\big| (\mathcal{A}_{\mathbf{q}, \mathbf{k}})_{\mu \mu'} \big|^2 \,
\delta (\omega - \varepsilon_{\mathbf{k}, \mu} -
\varepsilon_{\mathbf{q} - \mathbf{k}, \mu'})
\nonumber \\
&& \times \, \Theta (\varepsilon_{\mathbf{k}, \mu}) \, \Theta
(\varepsilon_{\mathbf{q} - \mathbf{k}, \mu'}), \label{eq-I}
\end{eqnarray}
where the individual amplitudes $(\mathcal{A}_{\mathbf{q},
\mathbf{k}})_{\mu \mu'}$ are derived in the SM \cite{SM} to be
appropriate matrix elements of
\begin{equation}
\mathcal{A}_{\mathbf{q}, \mathbf{k}}^{\phantom{\dag}} =
\mathcal{E}_{\mathbf{k}}^{\phantom{\dag}} \cdot
\mathcal{W}_{\mathbf{k}}^{\dag} \cdot \mathcal{W}_{\mathbf{q} -
\mathbf{k}}^{*} - \mathcal{W}_{\mathbf{k}}^{\dag} \cdot
\mathcal{W}_{\mathbf{q} - \mathbf{k}}^{*} \cdot
\mathcal{E}_{\mathbf{q} - \mathbf{k}}^{\phantom{\dag}}.
\label{eq-A-3}
\end{equation}
From a computational point of view, the intensity $I_0 (\omega,
\mathbf{q})$ is obtained numerically as a histogram of
$|(\mathcal{A}_{\mathbf{q}, \mathbf{k}})_{\mu \mu'}|^2$ in terms of
the final-state energies $\omega = \varepsilon_{\mathbf{k}, \mu} +
\varepsilon_{\mathbf{q} - \mathbf{k}, \mu'}$.

\begin{figure}
\includegraphics[width=0.94\columnwidth]{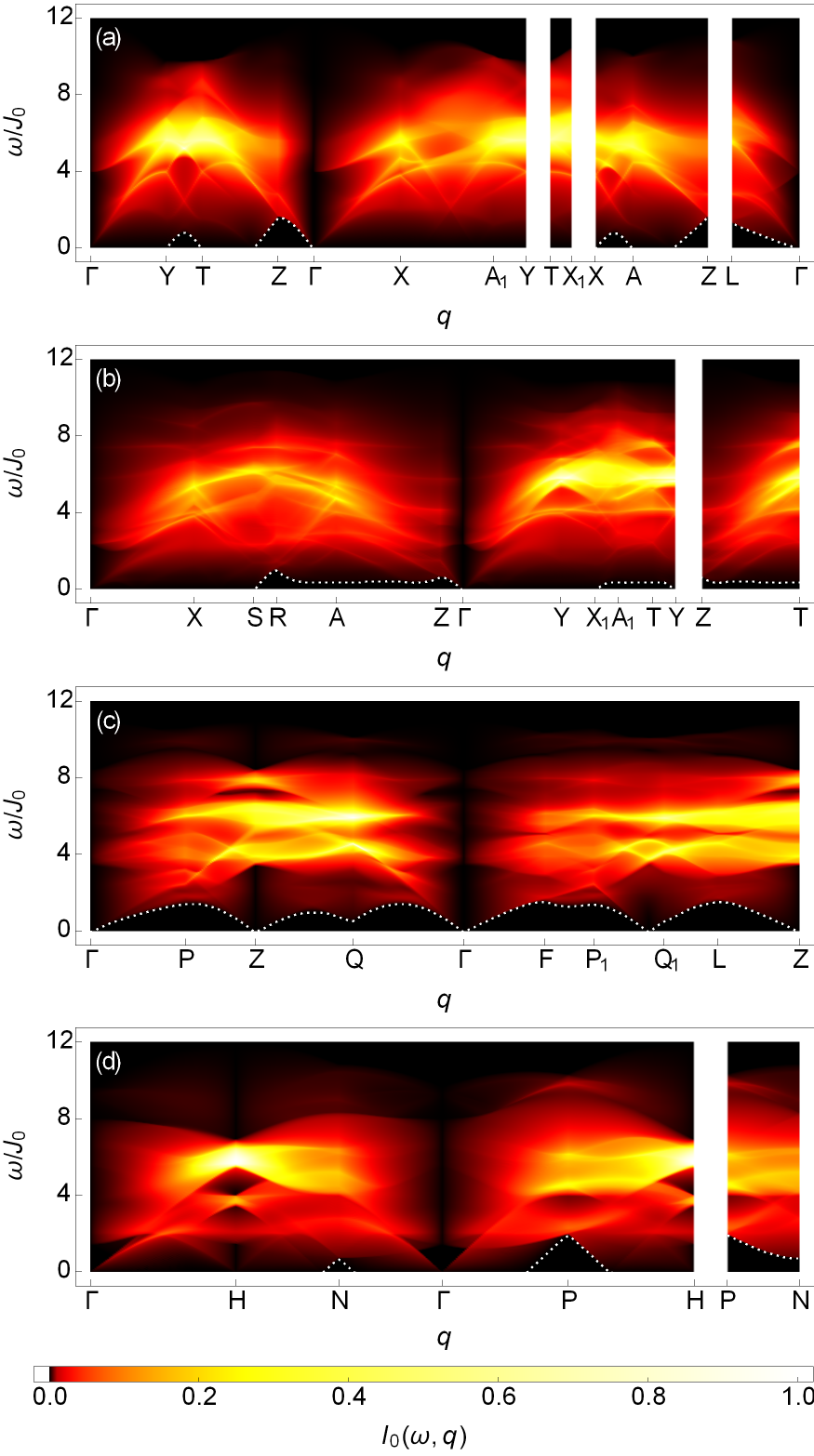}
\caption{Lowest-order SC RIXS intensities of isotropic Kitaev models
($J_{x,y,z} = J_0$) on the (a) hyperhoneycomb, (b) stripyhoneycomb,
(c) hyperhexagon, and (d) hyperoctagon lattices. In each case, the
intensity is plotted along the high-symmetry path depicted in
Fig.~\ref{fig-3} and is normalized to be between $0$ and $1$. The
dotted white line indicates a gap, below which the intensity is
exactly zero.} \label{fig-2}
\end{figure}

\begin{figure}
\includegraphics[width=0.94\columnwidth]{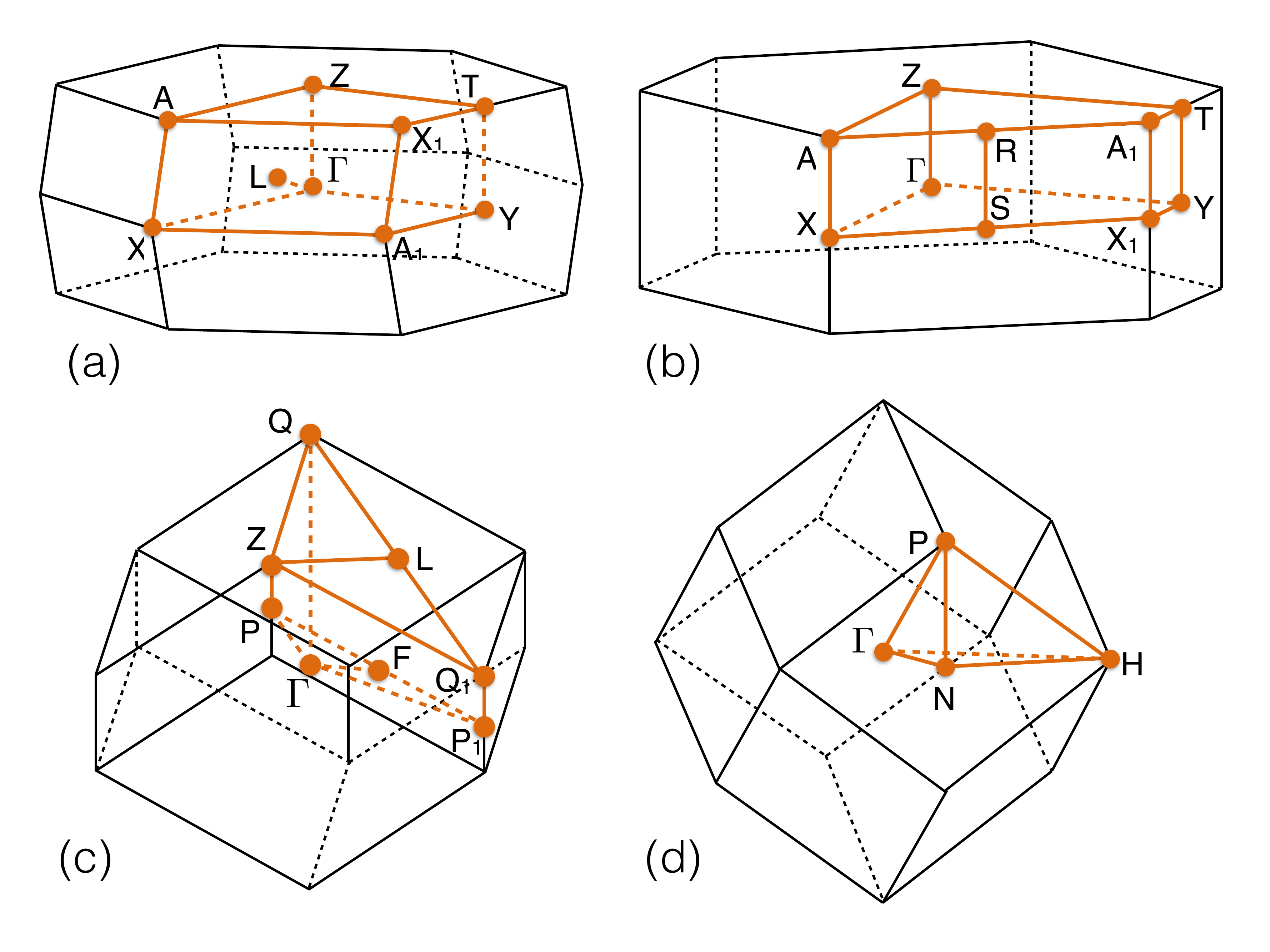}
\caption{High-symmetry paths \cite{Setyawan-2010} within the
Brillouin zones of the (a) hyperhoneycomb, (b) stripyhoneycomb, (c)
hyperhexagon, and (d) hyperoctagon lattices.} \label{fig-3}
\end{figure}

\emph{Results and discussion.---}At the isotropic point of each
Kitaev model, the lowest-order SC RIXS response $I_0 (\omega,
\mathbf{q})$ is plotted in Fig.~\ref{fig-2} along a high-symmetry
path \cite{Setyawan-2010} within the Brillouin zone depicted in
Fig.~\ref{fig-3}. For each model, the lack of sharp dispersion
curves $\omega (\mathbf{q})$ indicates the absence of a one-fermion
response, which is forbidden due to the fractionalized nature of the
fermions. Instead, the SC RIXS response in the experimental regime
is dominated by the two-fermion response in Eq.~(\ref{eq-I}), and
the overall energy dependence of each response is thus proportional
to the two-fermion joint density of states plotted in the SM
\cite{SM}. Since the fermion bandwidth is $\approx 6 J_0$, the
bandwidth of the response is then $\approx 12 J_0$.

Unlike the INS responses \cite{Smith-2015, Smith-2016} or,
equivalently, the NSC RIXS responses, the SC RIXS responses in
Fig.~\ref{fig-2} are gapless and they have a pronounced momentum
dependence. For each model, the low-energy (gapless) response is
determined by the nodal structure of the fermions. Since the
lowest-order SC RIXS processes create two fermions, the
corresponding response is gapless at momentum $\mathbf{q}$ if there
are gapless fermions at some momenta $\mathbf{k}_1$ and
$\mathbf{k}_2$ such that $\mathbf{q} = \mathbf{k}_1 + \mathbf{k}_2$.
For the hyperhexagon model, the fermions are gapless at Weyl points,
and the response is thus only gapless at particular points of the
Brillouin zone. For the hyperhoneycomb and the stripyhoneycomb
models, the fermions are gapless along a nodal line within the
$\Gamma$-X-Y plane, and the response is thus gapless in most of the
$\Gamma$-X-Y plane for both models and also in most of the Z-A-T
plane for the hyperhoneycomb model. However, it is still gapped at a
generic point of the Brillouin zone between these high-symmetry
planes. For the hyperoctagon model, the fermions are gapless on a
Fermi surface, and the response is thus gapless in most of the
Brillouin zone.

For each model, the SC RIXS response in Fig.~\ref{fig-2} is strongly
suppressed around the $\Gamma$ point. Indeed, since
$\mathcal{E}_{\mathbf{k}}^{\phantom{\dag}} =
-\mathcal{E}_{-\mathbf{k}}^{\phantom{\dag}}$ is diagonal and
$\mathcal{W}_{\mathbf{k}}^{\phantom{\dag}} =
\mathcal{W}_{-\mathbf{k}}^{*}$ is unitary,
$(\mathcal{W}_{\mathbf{k}}^{\dag} \cdot \mathcal{W}_{\mathbf{q} -
\mathbf{k}}^{*})_{\mu \mu'} = \delta_{\mu \mu'}$ and hence
$\mathcal{A}_{\mathbf{q}, \mathbf{k}}$ is purely diagonal for
$\mathbf{q} = \mathbf{0}$. The intensity $I_0 (\omega, \mathbf{0})$
in Eq.~(\ref{eq-I}) is then zero due to the Heaviside step functions
and $\varepsilon_{-\mathbf{k}, \mu} = -\varepsilon_{\mathbf{k},
\mu}$. From a physical point of view, this suppression of the
intensity can be understood for each model as a destructive
interference between scattering processes at the two sublattices of
the bipartite lattice, which in turn arises because each scattering
process creates two fermions and each fermion involves a phase
factor $i$ between the two sublattices (see the SM \cite{SM}).
Remarkably, the phase factor $i$ indicates that the appropriate
symmetry exchanging the two sublattices \cite{Footnote-2} acts
projectively on the fermions as its action on them squares to $-1$
instead of $+1$ \cite{You-2012}. The strong suppression of the
response around the $\Gamma$ point is thus a further signature of
(symmetry) fractionalization.

For any actual material realizing a KSL phase, the Hamiltonian
necessarily contains additional terms with respect to those in
Eq.~(\ref{eq-H-1}). In general, the high-energy response is robust
against such perturbations, even beyond the phase transition into an
ordered phase \cite{Banerjee-2016}, but the low-energy response of a
generic KSL can be completely different from that of a pure Kitaev
model \cite{Song-2016}. Nevertheless, we expect that the low-energy
features of each SC RIXS response in Fig.~\ref{fig-2} are valid for
a generic point of the corresponding KSL phase as the low-energy
physics is still governed by gapless (dressed) fermions with a
particular nodal structure protected by the (projective) symmetries
of the system \cite{Hermanns-2014, Hermanns-2015, O'Brien-2016}. In
particular, for the hyperhoneycomb and the stripyhoneycomb KSLs, the
nodal line remains within the $\Gamma$-X-Y plane as long as the
two-fold rotation symmetry around any $z$ bond is intact
\cite{Footnote-3}. The suppression of the response around the
$\Gamma$ point is also expected to be a robust feature of each KSL
phase as it occurs due to the particular way the symmetries
fractionalize when acting on the fermions. In fact, it should be
present for any KSL on a bipartite lattice, including the honeycomb
KSL \cite{Halasz-2016}.

\emph{Summary.---}Calculating the exact RIXS responses of four 3D
Kitaev models, we have demonstrated that RIXS is a sensitive probe
of the fractionalized excitations in 3D KSLs. In its NSC channels,
RIXS measures the dynamical spin structure factor, while in its SC
channel, it gives a complementary response, picking up exclusively
the Majorana fermions. By looking at where the SC RIXS response is
gapless, one can distinguish between the various nodal structures of
Majorana fermions possible in 3D KSLs. Conversely, the suppression
of the response around the $\Gamma$ point is expected to be a
generic signature of all KSLs on a bipartite lattice.

We thank J.~van den Brink, F.~J.~Burnell, J.~T.~Chalker, and
J.~Knolle for collaboration on closely related topics. G.~B.~H. is
supported by the Gordon and Betty Moore Foundation's EPiQS
Initiative through Grant No.~GBMF4304. N.~B.~P. is supported by the
NSF Grant No.~DMR-1511768 and is also grateful to the Perimeter
Institute for their hospitality during the course of this work.
Research at the Perimeter Institute is supported by the Government
of Canada through Industry Canada and by the Province of Ontario
through the Ministry of Economic Development and Innovation.

%%%%%%%%%%%%%%%%%%%%%%%%%%%%%%%%%%%%%%%%%%%%%%%%%

%%%%%%%%%%%%%%%%%%%%%%%%%%%%%%%%%%%%%%%%%%%%%%%%%

\clearpage

\begin{widetext}

\subsection{\large Supplementary Material}

\section{Tricoordinated lattices and Majorana-fermion Hamiltonians} \label{sec-latt}

Here we provide additional information on the four tricoordinated
three-dimensional (3D) lattices as well as the corresponding
Majorana-fermion Hamiltonians introduced in the main text. For each
lattice, the unit cell and the lattice vectors are depicted in
Fig.~\ref{fig-S1}, the corresponding Majorana-fermion band structure
is plotted in Fig.~\ref{fig-S2}, while the one-fermion density of
states and the two-fermion (joint) density of states are plotted in
Figs.~\ref{fig-S3} and \ref{fig-S4}. In addition, the formal
descriptions of the individual lattices and the precise forms of the
corresponding Hamiltonian matrices $\hat{\mathcal{H}}_{\mathbf{k}}$
are given below. Note that we describe each lattice in an orthogonal
coordinate system where the distance between two neighboring sites
is assumed to be unity.

\subsection{Hyperhoneycomb lattice} \label{sec-latt-A}

The hyperhoneycomb lattice is a face-centered orthorhombic lattice
with four sites per (primitive) unit cell. The three lattice vectors
of the face-centered orthorhombic lattice are given by
\begin{eqnarray}
& \mathbf{a}_1 = \left( 0, \sqrt{2}, 3 \right), \quad \mathbf{a}_2 =
\left( 1, 0, 3 \right), \quad \mathbf{a}_3 = \left( 1, \sqrt{2}, 0
\right), \label{eq-latt-A-a}
\end{eqnarray}
while the coordinates of the four sites in each unit cell are
\begin{eqnarray}
& \mathbf{r}_1 = \left( 0, 0, 0 \right), \quad \mathbf{r}_2 = \left(
\frac{1}{2}, \frac{\sqrt{2}} {2}, \frac{3}{2} \right), \quad
\mathbf{r}_3 = \left( 0, 0, 1 \right), \quad \mathbf{r}_4 = \left(
\frac{1}{2}, \frac{\sqrt{2}} {2}, \frac{5}{2} \right).
\label{eq-latt-A-r}
\end{eqnarray}
In the notation of Ref.~[55] in the main text, the Brillouin zone is
of type ORCF$_1$, and its high-symmetry points have coordinates
\begin{eqnarray}
& \Gamma = \left( 0, 0, 0 \right), \quad \textrm{A} = \left(
\frac{25 \pi} {36}, 0, \frac{\pi} {3} \right), \quad \textrm{A}_1 =
\left( \frac{11 \pi} {36}, \frac{\pi} {\sqrt{2}}, 0 \right), \quad
\textrm{L} = \left( \frac{\pi} {2}, \frac{\pi} {2 \sqrt{2}},
\frac{\pi} {6} \right), \quad \textrm{T} = \left( 0, \frac{\pi}
{\sqrt{2}}, \frac{\pi} {3} \right), \nonumber \\
& \textrm{X} = \left( \frac{29 \pi} {36}, 0, 0 \right), \quad
\textrm{X}_1 = \left( \frac{7 \pi} {36}, \frac{\pi} {\sqrt{2}},
\frac{\pi} {3} \right), \quad \textrm{Y} = \left( 0, \frac{\pi}
{\sqrt{2}}, 0 \right), \quad \textrm{Z} = \left( 0, 0, \frac{\pi}
{3} \right). \label{eq-latt-A-BZ}
\end{eqnarray}
In the flux-free sector of the corresponding Kitaev model, the
Hamiltonian matrix $\hat{\mathcal{H}}_{\mathbf{k}}$ takes the form
\begin{equation}
\hat{\mathcal{H}}_{\mathbf{k}}^{\phantom{\dag}} =
\Omega_{\mathbf{k}}^{\dag} \cdot
\check{\mathcal{H}}_{\mathbf{k}}^{\phantom{\dag}} \cdot
\Omega_{\mathbf{k}}^{\phantom{\dag}}, \qquad
\check{\mathcal{H}}_{\mathbf{k}}^{\phantom{\dag}} = \left(
\begin{array}{cc} 0 & -i
\check{\mathcal{M}}_{\mathbf{k}}^{\phantom{\dag}} \\ i
\check{\mathcal{M}}_{\mathbf{k}}^{\dag} & 0
\end{array} \right), \qquad
\check{\mathcal{M}}_{\mathbf{k}}^{\phantom{\dag}} = \left(
\begin{array}{cc} J_z & J_x e^{-i k_2} + J_y e^{-i k_1} \\ J_x + J_y
e^{i k_3} & J_z \end{array} \right), \label{eq-latt-A-H}
\end{equation}
where $k_j \equiv \mathbf{k} \cdot \mathbf{a}_j$ with $j = 1,2,3$,
while $\Omega_{\mathbf{k}} = \mathrm{diag} \{ \exp (i \mathbf{k}
\cdot \mathbf{r}_j) \}_{j = 1,2,3,4}$.

\subsection{Stripyhoneycomb lattice} \label{sec-latt-B}

The stripyhoneycomb lattice is a base-centered orthorhombic lattice
with eight sites per (primitive) unit cell. The three lattice
vectors of the base-centered orthorhombic lattice are given by
\begin{eqnarray}
& \mathbf{a}_1 = \left( 1, -\sqrt{2}, 0 \right), \quad \mathbf{a}_2
= \left( 1, \sqrt{2}, 0 \right), \quad \mathbf{a}_3 = \left( 0, 0, 6
\right), \label{eq-latt-B-a}
\end{eqnarray}
while the coordinates of the eight sites in each unit cell are
\begin{eqnarray}
& \mathbf{r}_1 = \left( 0, 0, 0 \right), \quad \mathbf{r}_2 = \left(
\frac{1}{2}, \frac{\sqrt{2}} {2}, \frac{3}{2} \right), \quad
\mathbf{r}_3 = \left( 0, 0, 3 \right), \quad \mathbf{r}_4 = \left(
\frac{1}{2}, -\frac{\sqrt{2}} {2}, \frac{9}{2} \right), \nonumber \\
& \mathbf{r}_5 = \left( 0, 0, 1 \right), \quad \mathbf{r}_6 = \left(
\frac{1}{2}, \frac{\sqrt{2}} {2}, \frac{5}{2} \right), \quad
\mathbf{r}_7 = \left( 0, 0, 4 \right), \quad \mathbf{r}_8 = \left(
\frac{1}{2}, -\frac{\sqrt{2}} {2}, \frac{11}{2} \right).
\label{eq-latt-B-r}
\end{eqnarray}
In the notation of Ref.~[55] in the main text, the Brillouin zone is
of type ORCC, and its high-symmetry points have coordinates
\begin{eqnarray}
& \Gamma = \left( 0, 0, 0 \right), \quad \textrm{A} = \left( \frac{3
\pi} {4}, 0, \frac{\pi} {6} \right), \quad \textrm{A}_1 = \left(
\frac{\pi} {4}, \frac{\pi} {\sqrt{2}}, \frac{\pi} {6} \right), \quad
\textrm{R} = \left( \frac{\pi} {2}, \frac{\pi} {2 \sqrt{2}},
\frac{\pi} {6} \right), \quad \textrm{S} = \left( \frac{\pi} {2},
\frac{\pi} {2 \sqrt{2}}, 0 \right), \quad \textrm{T} = \left( 0,
\frac{\pi} {\sqrt{2}}, \frac{\pi} {6} \right),
\nonumber \\
& \textrm{X} = \left( \frac{3 \pi} {4}, 0, 0 \right), \quad
\textrm{X}_1 = \left( \frac{\pi} {4}, \frac{\pi} {\sqrt{2}}, 0
\right), \quad \textrm{Y} = \left( 0, \frac{\pi} {\sqrt{2}}, 0
\right), \quad \textrm{Z} = \left( 0, 0, \frac{\pi} {6} \right).
\label{eq-latt-B-BZ}
\end{eqnarray}
In the flux-free sector of the corresponding Kitaev model, the
Hamiltonian matrix $\hat{\mathcal{H}}_{\mathbf{k}}$ takes the form
\begin{eqnarray}
&&\qquad \qquad \qquad
\hat{\mathcal{H}}_{\mathbf{k}}^{\phantom{\dag}} =
\Omega_{\mathbf{k}}^{\dag} \cdot
\check{\mathcal{H}}_{\mathbf{k}}^{\phantom{\dag}} \cdot
\Omega_{\mathbf{k}}^{\phantom{\dag}}, \qquad
\check{\mathcal{H}}_{\mathbf{k}}^{\phantom{\dag}} = \left(
\begin{array}{cc} 0 & -i
\check{\mathcal{M}}_{\mathbf{k}}^{\phantom{\dag}} \\ i
\check{\mathcal{M}}_{\mathbf{k}}^{\dag} & 0
\end{array} \right), \nonumber \\
\nonumber \\
&& \check{\mathcal{M}}_{\mathbf{k}}^{\phantom{\dag}} = \left(
\begin{array}{cccc} J_z & 0 & 0 & J_x e^{-i k_1 - i k_3} + J_y
e^{-i k_3} \\ J_x + J_y e^{i k_2} & J_z & 0 & 0 \\ 0 & J_x e^{-i
k_2} + J_y & J_z & 0 \\ 0 & 0 & J_x + J_y e^{i k_1} & J_z
\end{array} \right), \label{eq-latt-B-H}
\end{eqnarray}
where $k_j \equiv \mathbf{k} \cdot \mathbf{a}_j$ with $j = 1,2,3$,
while $\Omega_{\mathbf{k}} = \mathrm{diag} \{ \exp (i \mathbf{k}
\cdot \mathbf{r}_j) \}_{j = 1, 2, \ldots, 8}$.

\subsection{Hyperhexagon lattice} \label{sec-latt-C}

The hyperhexagon lattice is a simple rhombohedral lattice with six
sites per (primitive) unit cell. The three lattice vectors of the
rhombohedral lattice are given by
\begin{eqnarray}
& \mathbf{a}_1 = \left( 0, -\frac{5} {\sqrt{3}}, \sqrt{\frac{2}{3}}
\right), \quad \mathbf{a}_2 = \left( -\frac{5}{2}, \frac{5} {2
\sqrt{3}}, \sqrt{\frac{2}{3}} \right), \quad \mathbf{a}_3 = \left(
\frac{5}{2}, \frac{5} {2 \sqrt{3}}, \sqrt{\frac{2}{3}} \right),
\label{eq-latt-C-a}
\end{eqnarray}
while the coordinates of the six sites in each unit cell are
\begin{eqnarray}
& \mathbf{r}_1 = \left( 0, 0, 0 \right), \quad \mathbf{r}_2 = \left(
\frac{1}{2}, \frac{1} {2 \sqrt{3}}, 2 \sqrt{\frac{2}{3}} \right),
\quad \mathbf{r}_3 = \left( 1, \sqrt{3}, \sqrt{6} \right), \quad
\mathbf{r}_4 = \left( \frac{1}{2}, -\frac{1} {2 \sqrt{3}},
\sqrt{\frac{2}{3}} \right), \nonumber \\
& \mathbf{r}_5 = \left( 1, \frac{2} {\sqrt{3}}, 2 \sqrt{\frac{2}{3}}
\right), \quad \mathbf{r}_6 = \left( \frac{3}{2}, \frac{5} {2
\sqrt{3}}, 4 \sqrt{\frac{2}{3}} \right). \label{eq-latt-C-r}
\end{eqnarray}
In the notation of Ref.~[55] in the main text, the Brillouin zone is
of type RHL$_2$, and its high-symmetry points have coordinates
\begin{eqnarray}
& \Gamma = \left( 0, 0, 0 \right), \quad \textrm{F} = \left(
\frac{\pi} {5}, -\frac{\sqrt{3} \pi} {5}, 0 \right), \quad
\textrm{L} = \left( 0, -\frac{2 \pi} {5 \sqrt{3}}, \frac{\pi}
{\sqrt{6}} \right), \quad \textrm{P} = \left( \frac{2 \pi} {5},
-\frac{2 \pi} {5 \sqrt{3}}, \frac{2 \pi} {25} \sqrt{\frac{2}{3}}
\right), \quad \textrm{P}_1 = \left( 0, -\frac{4 \pi} {5 \sqrt{3}},
-\frac{2 \pi} {25} \sqrt{\frac{2}{3}} \right), \nonumber \\
& \textrm{Q} = \left( 0, 0, \frac{11 \pi} {25} \sqrt{\frac{3}{2}}
\right), \quad \textrm{Q}_1 = \left( 0, -\frac{4 \pi} {5 \sqrt{3}},
\frac{17 \pi} {25 \sqrt{6}} \right), \quad \textrm{Z} = \left(
\frac{2 \pi} {5},  -\frac{2 \pi} {5 \sqrt{3}}, \frac{\pi} {\sqrt{6}}
\right). \label{eq-latt-C-BZ}
\end{eqnarray}
In the ground-state flux sector of the corresponding Kitaev model,
the Hamiltonian matrix $\hat{\mathcal{H}}_{\mathbf{k}}$ takes the
form
\begin{equation}
\qquad \qquad \hat{\mathcal{H}}_{\mathbf{k}}^{\phantom{\dag}} =
\Omega_{\mathbf{k}}^{\dag} \cdot
\check{\mathcal{H}}_{\mathbf{k}}^{\phantom{\dag}} \cdot
\Omega_{\mathbf{k}}^{\phantom{\dag}}, \label{eq-latt-C-H} \\
\end{equation}
\begin{eqnarray}
\check{\mathcal{H}}_{\mathbf{k}}^{\phantom{\dag}} = i \left(
\begin{array}{cccccc} 0 & J_x e^{-i k_1 - i k_2 - i k_3} & 0 & J_z &
0 & J_y e^{-i k_1 - i k_2 - 2 i k_3} \\ -J_x e^{i k_1 + i k_2 + i
k_3} & 0 & 0 & J_y & J_z & 0 \\ 0 & 0 & 0 & J_x e^{i k_2 + i k_3} &
J_y & J_z \\ -J_z & -J_y & -J_x e^{-i k_2 - i k_3} & 0 & 0 & 0 \\ 0
& -J_z & -J_y & 0 & 0 & -J_x e^{-i k_1 - i k_2 - i k_3} \\ -J_y e^{i
k_1 + i k_2 + 2 i k_3} & 0 & -J_z & 0 & J_x e^{i k_1 + i k_2 + i
k_3} & 0 \end{array} \right), \nonumber
\end{eqnarray}
where $k_j \equiv \mathbf{k} \cdot \mathbf{a}_j$ with $j = 1,2,3$,
while $\Omega_{\mathbf{k}} = \mathrm{diag} \{ \exp (i \mathbf{k}
\cdot \mathbf{r}_j) \}_{j = 1, 2, \ldots, 6}$.

\begin{figure}
\includegraphics[width=1.0\columnwidth]{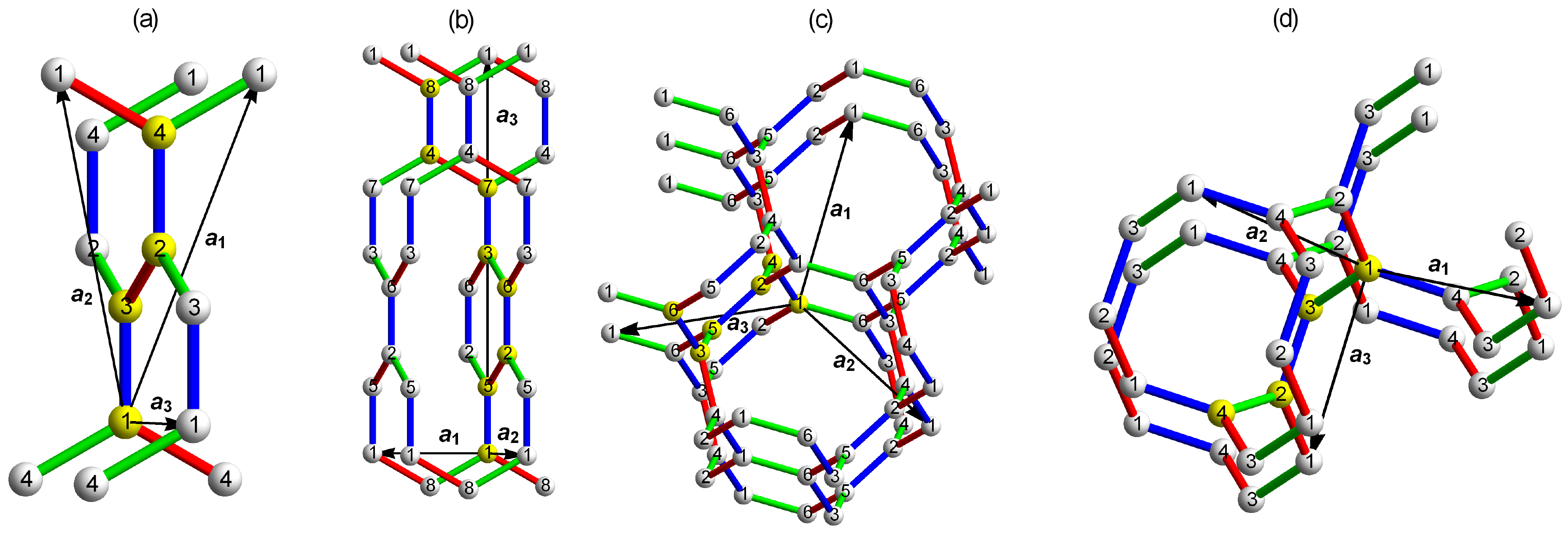}
\caption{Primitive unit cells (yellow sites) and lattice vectors
($\mathbf{a}_{1,2,3}$) of the (a) hyperhoneycomb, (b)
stripyhoneycomb, (c) hyperhexagon, and (d) hyperoctagon lattices.
Different bond types $x$, $y$, and $z$ are marked by red, green, and
blue, respectively.} \label{fig-S1}
\end{figure}

\begin{figure}
\includegraphics[width=0.9\columnwidth]{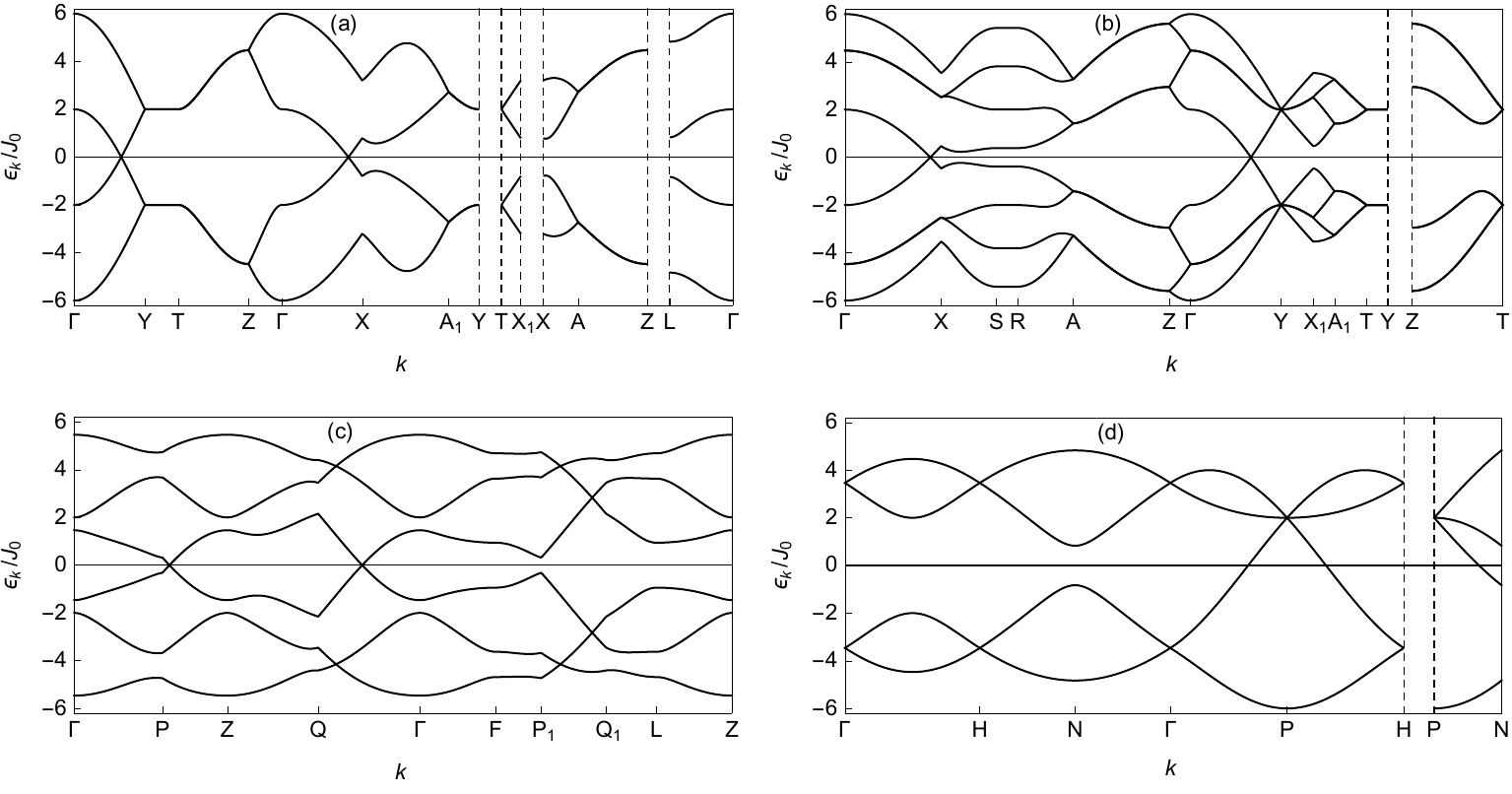}
\caption{Majorana-fermion (spinon) band structures of isotropic
Kitaev models ($J_{x,y,z} = J_0$) on the (a) hyperhoneycomb, (b)
stripyhoneycomb, (c) hyperhexagon, and (d) hyperoctagon lattices.
Note that only the fermions with positive energies
$\varepsilon_{\mathbf{k}} > 0$ are physical while the ones with
negative energies $\varepsilon_{\mathbf{k}} < 0$ at momentum
$\mathbf{k}$ are identified with physical fermions at momentum
$-\mathbf{k}$.} \label{fig-S2}
\end{figure}

\begin{figure}
\includegraphics[width=1.0\columnwidth]{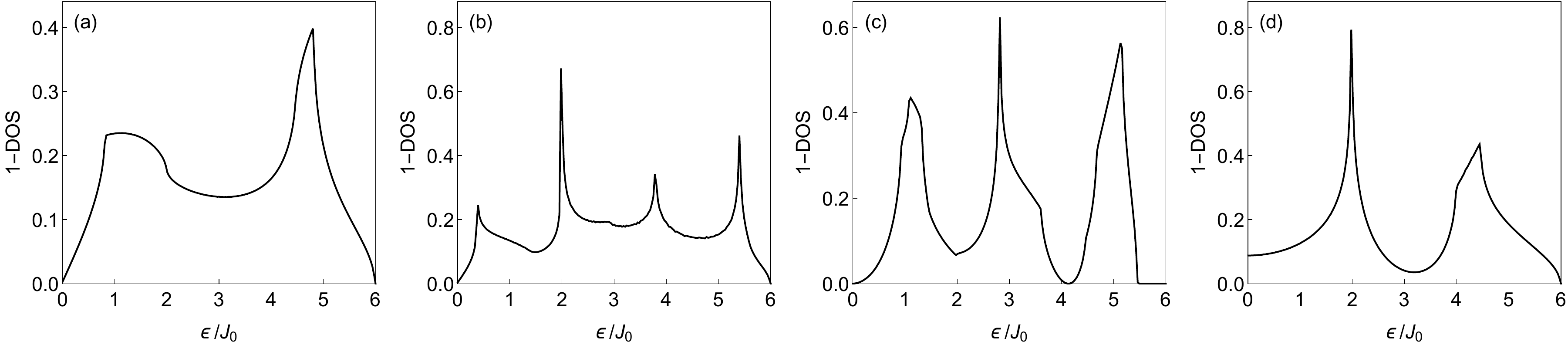}
\caption{One-fermion densities of states (1-DOS) of isotropic Kitaev
models ($J_{x,y,z} = J_0$) on the (a) hyperhoneycomb, (b)
stripyhoneycomb, (c) hyperhexagon, and (d) hyperoctagon lattices. In
each case, the density of states is normalized such that its
integral is unity.} \label{fig-S3}
\end{figure}

\begin{figure}
\includegraphics[width=1.0\columnwidth]{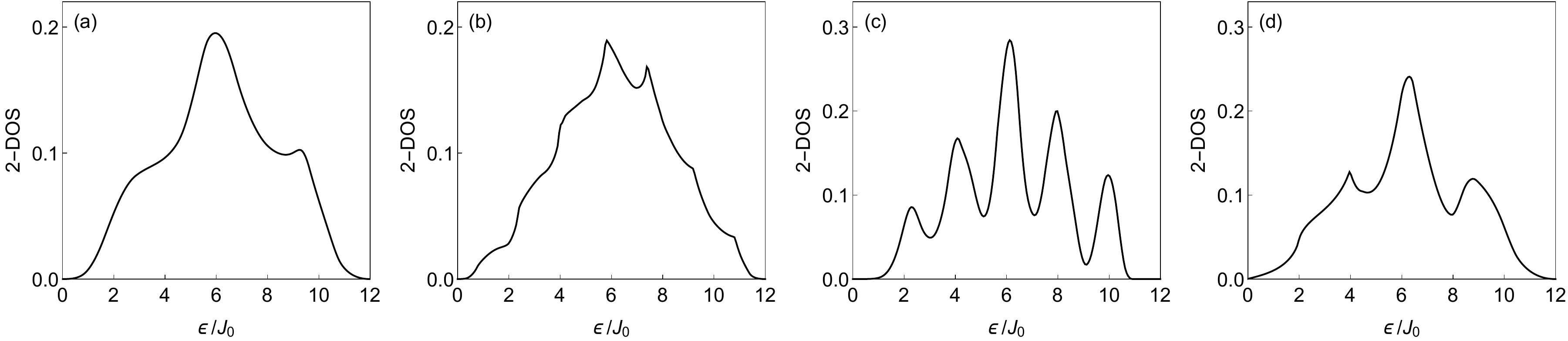}
\caption{Two-fermion densities of states (2-DOS) of isotropic Kitaev
models ($J_{x,y,z} = J_0$) on the (a) hyperhoneycomb, (b)
stripyhoneycomb, (c) hyperhexagon, and (d) hyperoctagon lattices. In
each case, the density of states is normalized such that its
integral is unity.} \label{fig-S4}
\end{figure}

\subsection{Hyperoctagon lattice} \label{sec-latt-D}

The hyperoctagon lattice is a body-centered cubic lattice with four
sites per (primitive) unit cell. The three lattice vectors of the
body-centered cubic lattice are given by
\begin{eqnarray}
& \mathbf{a}_1 = \left( -\sqrt{2}, \sqrt{2}, \sqrt{2} \right), \quad
\mathbf{a}_2 = \left( \sqrt{2}, -\sqrt{2}, \sqrt{2} \right), \quad
\mathbf{a}_3 = \left( \sqrt{2}, \sqrt{2}, -\sqrt{2} \right),
\label{eq-latt-D-a}
\end{eqnarray}
while the coordinates of the four sites in each unit cell are
\begin{eqnarray}
& \mathbf{r}_1 = \left( 0, 0, 0 \right), \quad \mathbf{r}_2 = \left(
\sqrt{2}, \frac{1} {\sqrt{2}}, -\frac{1} {\sqrt{2}} \right), \quad
\mathbf{r}_3 = \left( \frac{1} {\sqrt{2}}, 0, -\frac{1} {\sqrt{2}}
\right), \quad \mathbf{r}_4 = \left( \frac{3} {\sqrt{2}}, \frac{1}
{\sqrt{2}}, 0 \right). \label{eq-latt-D-r}
\end{eqnarray}
In the notation of Ref.~[55] in the main text, the Brillouin zone is
of type BCC, and its high-symmetry points have coordinates
\begin{equation}
\Gamma = \left( 0, 0, 0 \right), \quad \textrm{H} = \left( 0,
\frac{\pi} {\sqrt{2}}, 0 \right), \quad \textrm{N} = \left(
\frac{\pi} {2 \sqrt{2}}, \frac{\pi} {2 \sqrt{2}}, 0 \right), \quad
\textrm{P} = \left( \frac{\pi} {2 \sqrt{2}}, \frac{\pi} {2
\sqrt{2}}, \frac{\pi} {2 \sqrt{2}} \right). \label{eq-latt-D-BZ}
\end{equation}
In the flux-free sector of the corresponding Kitaev model, the
Hamiltonian matrix $\hat{\mathcal{H}}_{\mathbf{k}}$ takes the form
\begin{equation}
\hat{\mathcal{H}}_{\mathbf{k}}^{\phantom{\dag}} =
\Omega_{\mathbf{k}}^{\dag} \cdot
\check{\mathcal{H}}_{\mathbf{k}}^{\phantom{\dag}} \cdot
\Omega_{\mathbf{k}}^{\phantom{\dag}}, \qquad
\check{\mathcal{H}}_{\mathbf{k}}^{\phantom{\dag}} = i \left(
\begin{array}{cccc} 0 & -J_x e^{-i k_3} & -J_y & -J_z e^{-i k_2 - i
k_3} \\ J_x e^{i k_3} & 0 & -J_z & J_y \\ J_y & J_z & 0 & -J_x e^{-i
k_1 - i k_2 - i k_3} \\ J_z e^{i k_2 + i k_3} & -J_y & J_x e^{i k_1
+ i k_2 + i k_3} & 0 \end{array} \right), \label{eq-latt-D-H}
\end{equation}
where $k_j \equiv \mathbf{k} \cdot \mathbf{a}_j$ with $j = 1,2,3$,
while $\Omega_{\mathbf{k}} = \mathrm{diag} \{ \exp (i \mathbf{k}
\cdot \mathbf{r}_j) \}_{j = 1,2,3,4}$.

\section{Scattering amplitude in the spin-conserving channel} \label{sec-amp}

Here we evaluate the spin-conserving RIXS amplitude $A_0 (m,
\mathbf{q})$ in Eq.~(7) of the main text between the ground state $|
0 \rangle$ and a generic final state $| m \rangle = \psi_{\mathbf{q}
- \mathbf{k}, \mu'}^{\dag} \psi_{\mathbf{k}, \mu}^{\dag} | 0
\rangle$ containing two fermions. The inverse of Eq.~(4) in the main
text is given by
\begin{equation}
c_{\mathbf{r} \in \nu}^{\phantom{\dag}} = \sqrt{\frac{2}{N}}
\sum_{\mathbf{k}, \mu} \left( \mathcal{W}_{\mathbf{k}}^{\dag}
\right)_{\mu \nu} \psi_{\mathbf{k}, \mu}^{\dag} \, e^{-i \mathbf{k}
\cdot \mathbf{r}}, \label{eq-amp-c}
\end{equation}
and the RIXS amplitude in Eq.~(7) of the main text reads
\begin{eqnarray}
A_0 (m, \mathbf{q}) &\propto& \sum_{\nu, \nu'} \, \sum_{\mathbf{r}
\in \nu} \, \sum_{\mathbf{r}' \in \nu'} e^{i \mathbf{q} \cdot
\mathbf{r}} \, \mathcal{H}_{\mathbf{r},
\mathbf{r}'}^{\phantom{\dag}} \langle 0 | \psi_{\mathbf{k},
\mu}^{\phantom{\dag}} \psi_{\mathbf{q} - \mathbf{k},
\mu'}^{\phantom{\dag}} c_{\mathbf{r}}^{\phantom{\dag}}
c_{\mathbf{r}'}^{\phantom{\dag}} | 0 \rangle
\nonumber \\
&=& \frac{2}{N} \sum_{\nu, \nu'} \, \sum_{\mathbf{r} \in \nu} \,
\sum_{\mathbf{r}' \in \nu'} e^{i \mathbf{q} \cdot \mathbf{r}} \,
\mathcal{H}_{\mathbf{r}, \mathbf{r}'}^{\phantom{\dag}} \bigg[
\langle 0 | \psi_{\mathbf{k}, \mu}^{\phantom{\dag}} \psi_{\mathbf{q}
- \mathbf{k}, \mu'}^{\phantom{\dag}} \psi_{\mathbf{k}, \mu}^{\dag}
\psi_{\mathbf{q} - \mathbf{k}, \mu'}^{\dag} | 0 \rangle \left(
\mathcal{W}_{\mathbf{k}}^{\dag} \right)_{\mu \nu} \left(
\mathcal{W}_{\mathbf{q} - \mathbf{k}}^{\dag} \right)_{\mu' \nu'}
e^{-i \mathbf{k} \cdot \mathbf{r} - i (\mathbf{q} - \mathbf{k})
\cdot \mathbf{r}'} \nonumber \\
&& + \, \langle 0 | \psi_{\mathbf{k}, \mu}^{\phantom{\dag}}
\psi_{\mathbf{q} - \mathbf{k}, \mu'}^{\phantom{\dag}}
\psi_{\mathbf{q} - \mathbf{k}, \mu'}^{\dag} \psi_{\mathbf{k},
\mu}^{\dag} | 0 \rangle \left( \mathcal{W}_{\mathbf{q} -
\mathbf{k}}^{\dag} \right)_{\mu' \nu} \left(
\mathcal{W}_{\mathbf{k}}^{\dag} \right)_{\mu \nu'} e^{-i (\mathbf{q}
- \mathbf{k}) \cdot \mathbf{r} - i \mathbf{k} \cdot \mathbf{r}'}
\bigg]. \label{eq-amp-A-1}
\end{eqnarray}
Using Eq.~(5) of the main text, the RIXS amplitude then becomes
\begin{eqnarray}
A_0 (m, \mathbf{q}) &\propto& \sum_{\nu, \nu'} \left[ \Big(
\hat{\mathcal{H}}_{\mathbf{k}}^{*} \Big)_{\nu \nu'} \left(
\mathcal{W}_{\mathbf{q} - \mathbf{k}}^{\dag} \right)_{\mu' \nu}
\left( \mathcal{W}_{\mathbf{k}}^{\dag} \right)_{\mu \nu'} - \Big(
\hat{\mathcal{H}}_{\mathbf{q} - \mathbf{k}}^{*} \Big)_{\nu \nu'}
\left( \mathcal{W}_{\mathbf{k}}^{\dag} \right)_{\mu \nu} \left(
\mathcal{W}_{\mathbf{q} - \mathbf{k}}^{\dag} \right)_{\mu' \nu'}
\right] \nonumber \\
&=& \left[ \mathcal{W}_{\mathbf{k}}^{\dag} \cdot
\hat{\mathcal{H}}_{\mathbf{k}}^{\phantom{\dag}} \cdot
\mathcal{W}_{\mathbf{q} - \mathbf{k}}^{*} -
\mathcal{W}_{\mathbf{k}}^{\dag} \cdot \hat{\mathcal{H}}_{\mathbf{q}
- \mathbf{k}}^{*} \cdot \mathcal{W}_{\mathbf{q} - \mathbf{k}}^{*}
\right]_{\mu \mu'}. \label{eq-amp-A-2}
\end{eqnarray}
Note that the relative minus sign between the two terms arises
because the two fermions $\psi_{\mathbf{k}, \mu}^{\dag}$ and
$\psi_{\mathbf{q} - \mathbf{k}, \mu'}^{\dag}$ are created in
opposite orders. Finally, due to
$\hat{\mathcal{H}}_{\mathbf{k}}^{\phantom{\dag}} =
\mathcal{W}_{\mathbf{k}}^{\phantom{\dag}} \cdot
\mathcal{E}_{\mathbf{k}}^{\phantom{\dag}} \cdot
\mathcal{W}_{\mathbf{k}}^{\dag}$ and
$\mathcal{W}_{\mathbf{k}}^{\phantom{\dag}} \cdot
\mathcal{W}_{\mathbf{k}}^{\dag} = 1$, the RIXS amplitude in
Eq.~(\ref{eq-amp-A-2}) takes the form
\begin{equation}
A_0 (m, \mathbf{q}) \propto \left[
\mathcal{E}_{\mathbf{k}}^{\phantom{\dag}} \cdot
\mathcal{W}_{\mathbf{k}}^{\dag} \cdot \mathcal{W}_{\mathbf{q} -
\mathbf{k}}^{*} - \mathcal{W}_{\mathbf{k}}^{\dag} \cdot
\mathcal{W}_{\mathbf{q} - \mathbf{k}}^{*} \cdot
\mathcal{E}_{\mathbf{q} - \mathbf{k}}^{\phantom{\dag}} \right]_{\mu
\mu'}. \label{eq-amp-A-3}
\end{equation}
This result is identical to the appropriate matrix element of
$\mathcal{A}_{\mathbf{q}, \mathbf{k}}$ in Eq.~(9) of the main text.

\section{Consequences of the bipartite lattice structure} \label{sec-sub}

For a Kitaev model on a bipartite lattice, where the $n = 2m$ sites
per unit cell can be divided into two classes $\nu_A = 1, \ldots, m$
and $\nu_B = m + 1, \ldots, 2m$ such that any $A$ ($B$) site only
neighbors $B$ ($A$) sites, the matrices
$\hat{\mathcal{H}}_{\mathbf{k}}^{\phantom{\dag}}$,
$\mathcal{E}_{\mathbf{k}}^{\phantom{\dag}}$, and
$\mathcal{W}_{\mathbf{k}}^{\phantom{\dag}}$ take the forms
\begin{equation}
\hat{\mathcal{H}}_{\mathbf{k}}^{\phantom{\dag}} = \left(
\begin{array}{cc} 0 & -i
\hat{\mathcal{M}}_{\mathbf{k}}^{\phantom{\dag}} \\
i \hat{\mathcal{M}}_{\mathbf{k}}^{\dag} & 0 \end{array} \right),
\quad \mathcal{E}_{\mathbf{k}}^{\phantom{\dag}} = \left(
\begin{array}{cc} \Lambda_{\mathbf{k}}^{\phantom{\dag}} & 0 \\
0 & -\Lambda_{\mathbf{k}}^{\phantom{\dag}} \end{array} \right),
\quad \mathcal{W}_{\mathbf{k}}^{\phantom{\dag}} = \frac{1}
{\sqrt{2}} \left( \begin{array}{cc}
\mathcal{U}_{\mathbf{k}}^{\phantom{\dag}} &
\mathcal{U}_{\mathbf{k}}^{\phantom{\dag}} \\
i \mathcal{V}_{\mathbf{k}}^{\phantom{\dag}} & -i
\mathcal{V}_{\mathbf{k}}^{\phantom{\dag}} \end{array} \right),
\label{eq-sub-H}
\end{equation}
where the diagonal matrix $\Lambda_{\mathbf{k}}^{\phantom{\dag}}$
and the unitary matrices $\mathcal{U}_{\mathbf{k}}^{\phantom{\dag}}$
and $\mathcal{V}_{\mathbf{k}}^{\phantom{\dag}}$ can be obtained by
taking the singular-value decomposition
$\hat{\mathcal{M}}_{\mathbf{k}}^{\phantom{\dag}} =
\mathcal{U}_{\mathbf{k}}^{\phantom{\dag}} \cdot
\Lambda_{\mathbf{k}}^{\phantom{\dag}} \cdot
\mathcal{V}_{\mathbf{k}}^{\dag}$. Since the singular values
$\lambda_{\mathbf{k}, \mu} = (\Lambda_{\mathbf{k}})_{\mu \mu}$ are
positive by definition, the physical (positive-energy) fermions are
then the ones with $\mu = 1, \ldots, m$ for all momenta
$\mathbf{k}$. Furthermore,
$\hat{\mathcal{M}}_{-\mathbf{k}}^{\phantom{\dag}} =
\hat{\mathcal{M}}_{\mathbf{k}}^{*}$ implies
$\Lambda_{-\mathbf{k}}^{\phantom{\dag}} =
\Lambda_{\mathbf{k}}^{\phantom{\dag}}$ as well as
$\mathcal{U}_{-\mathbf{k}}^{\phantom{\dag}} =
\mathcal{U}_{\mathbf{k}}^{*}$ and
$\mathcal{V}_{-\mathbf{k}}^{\phantom{\dag}} =
\mathcal{V}_{\mathbf{k}}^{*}$. We remark that all four Kitaev models
in the main text are defined on bipartite lattices. For the
hyperhoneycomb and the stripyhoneycomb models, the bipartite
structure of the lattice is compatible with the unit cell, and the
Hamiltonian matrices in Eqs.~(\ref{eq-latt-A-H}) and
(\ref{eq-latt-B-H}) readily take the form of
$\hat{\mathcal{H}}_{\mathbf{k}}^{\phantom{\dag}}$ in
Eq.~(\ref{eq-sub-H}). For the hyperhexagon and the hyperoctagon
models, the bipartite structure is not compatible with the original
unit cell of the lattice. However, if we artificially double the
unit cell, it becomes compatible with the bipartite structure, and
the enlarged Hamiltonian matrix takes the form of
$\hat{\mathcal{H}}_{\mathbf{k}}^{\phantom{\dag}}$ in
Eq.~(\ref{eq-sub-H}).

If the bipartite lattice of the Kitaev model consists of $N$ unit
cells, the physical (positive-energy) fermions with band index $\mu
= 1, \ldots, m$ in Eq.~(4) of the main text can be written as
\begin{equation}
\psi_{\mathbf{k}, \mu}^{\dag} = \frac{1} {2 \sqrt{N}} \sum_{\nu =
1}^{m} \left[ \big( \mathcal{U}_{\mathbf{k}}^{\phantom{\dag}}
\big)_{\nu \mu} \sum_{\mathbf{r} \in \nu_A}
c_{\mathbf{r}}^{\phantom{\dag}} \, e^{i \mathbf{k} \cdot \mathbf{r}}
+ i \big( \mathcal{V}_{\mathbf{k}}^{\phantom{\dag}} \big)_{\nu \mu}
\sum_{\mathbf{r} \in \nu_B} c_{\mathbf{r}}^{\phantom{\dag}} \, e^{i
\mathbf{k} \cdot \mathbf{r}} \right], \label{eq-sub-psi}
\end{equation}
where $\nu_A = \nu$ and $\nu_B = \nu + m$ for each $\nu = 1, \ldots,
m$. Restricting our attention to these physical fermions, the
spin-conserving RIXS amplitude in Eq.~(\ref{eq-amp-A-3}) is then
given by
\begin{equation}
A_0 (m, \mathbf{q}) \propto \left[
\Lambda_{\mathbf{k}}^{\phantom{\dag}} \cdot \left(
\mathcal{U}_{\mathbf{k}}^{\dag} \cdot \mathcal{U}_{\mathbf{q} -
\mathbf{k}}^{*} - \mathcal{V}_{\mathbf{k}}^{\dag} \cdot
\mathcal{V}_{\mathbf{q} - \mathbf{k}}^{*} \right) - \left(
\mathcal{U}_{\mathbf{k}}^{\dag} \cdot \mathcal{U}_{\mathbf{q} -
\mathbf{k}}^{*} - \mathcal{V}_{\mathbf{k}}^{\dag} \cdot
\mathcal{V}_{\mathbf{q} - \mathbf{k}}^{*} \right) \cdot
\Lambda_{\mathbf{q} - \mathbf{k}}^{\phantom{\dag}} \right]_{\mu
\mu'}. \label{eq-sub-A}
\end{equation}
The terms proportional to $\mathcal{U}_{\mathbf{k}}^{\dag} \cdot
\mathcal{U}_{\mathbf{q} - \mathbf{k}}^{*}$ capture scattering
processes at sublattice $A$ sites, while the terms proportional to
$\mathcal{V}_{\mathbf{k}}^{\dag} \cdot \mathcal{V}_{\mathbf{q} -
\mathbf{k}}^{*}$ capture scattering processes at sublattice $B$
sites. For $\mathbf{q} = \mathbf{0}$, the scattering processes in
each sublattice interfere constructively because
$\mathcal{U}_{\mathbf{k}}^{\dag} \cdot \mathcal{U}_{-\mathbf{k}}^{*}
= \mathcal{V}_{\mathbf{k}}^{\dag} \cdot
\mathcal{V}_{-\mathbf{k}}^{*} = 1$. However, there is a destructive
interference between the two sublattices due to the relative minus
sign in Eq.~(\ref{eq-sub-A}), and the spin-conserving RIXS intensity
$I_0 (\omega, \mathbf{q})$ in Eq.~(8) of the main text is thus zero.
Importantly, this relative minus sign between the two sublattices
arises because each scattering process creates two fermions and each
fermion involves a phase factor $i$ between the two sublattices [see
Eq.~(\ref{eq-sub-psi})].

\clearpage

\end{widetext}

%%%%%%%%%%%%%%%%%%%%%%%%%%%%%%%%%%%%%%%%%%%%%%%%%


\begin{references}
\bibitem{Footnote-1} In fact, the stripyhoneycomb model is
expected to have $\pi$ fluxes at certain elementary loops. However,
we have checked that this simplifying assumption does not affect our
main results.
\bibitem{Footnote-2} This symmetry is a two-fold rotation for the
hyperoctagon lattice and an inversion for the remaining three
lattices.
\bibitem{Footnote-3} If this symmetry is broken (say $J_x \neq J_y$),
the fermions are still gapless along a nodal line, and the response
is thus still gapless on a two-dimensional surface, but this surface
is no longer restricted to lie in a particular high-symmetry plane.
\end{references}
\end{document}